\documentclass{cup-hpl}

\usepackage{lipsum}
\usepackage{graphicx}
\usepackage[colorlinks=true,linkcolor=blue,citecolor=blue]{hyperref}%
\usepackage{epstopdf}
\usepackage{siunitx}

\begin{document}


\shorttitle{Positioning of Transparent Targets Using Defocusing Method in a Laser Proton Accelerator}
\shortauthor{Y. Shou et al.}

\title{Positioning of Transparent Targets Using Defocusing Method in a Laser Proton Accelerator}


\author[1]{Yinren Shou}
\author[1,2]{Dahui Wang}
\author[1]{Pengjie Wang}
\author[1]{Jianbo Liu}
\author[1]{Zhengxuan Cao}
\author[1]{Zhusong Mei}
\author[1]{Yixing Geng}
\author[1]{Jungao Zhu}
\author[1]{Qing Liao}
\author[1]{Yanying Zhao}
\author[1]{Chen Lin}
\author[1]{Haiyang Lu}
\author[1,3]{Wenjun Ma}
\author[1,3,4]{Xueqing Yan\corresp{W. Ma and X. Yan, State Key Laboratory of Nuclear Physics and Technology, Peking University, Beijing 100871, China. \email{wenjun.ma@pku.edu.cn, x.yan@pku.edu.cn}}}%

\address[1]{State Key Laboratory of Nuclear Physics and Technology, and Key Laboratory of HEDP of the Ministry of Education, CAPT, Peking University, Beijing 100871, China}
\address[2]{State Key Laboratory of Laser Interaction with Matter, Northwest Institute of Nuclear Technology, P. O. Box 69-26, Xi'an 710024, China}
\address[3]{Shenzhen Research Institute of Peking University, Shenzhen 518055, China}
\address[4]{Collaborative Innovation Center of Extreme Optics, Shanxi University, Taiyuan, Shanxi 030006, China}

\begin{abstract}
We report a positioning method for transparent targets with an accuracy of \SI{2}{\mu m} for a compact laser proton accelerator. The positioning system consists of two light-emitting diodes (LED), a long working distance objective and two charge coupled devices (CCD) for illumination, imaging and detection, respectively. We developed a defocusing method making transparent targets visible as phase objects and applied it to our system. Precise positioning of transparent targets can be realized by means of minimizing the image contrast of the phase objects. Fast positioning based on the relationship between the radius of spherical aberration ring and defocusing distance is also realized. Laser proton acceleration experiments have been performed to demonstrate the reliability of this positioning system.
\end{abstract}

\keywords{relativistic laser; Transparent targets positioning system; laser-driven ion source}

\maketitle
\section{Introduction}
Highly energetic ions can be generated when a relativistic laser pulse (intensity $I$ exceeding \SI{2.14e18}{W/cm^2} for wavelength $\lambda=$ \SI{800}{nm}) is focused on a solid target.\cite{tajima1979laser,strickland1985compression,macchi2013ion} Compact proton accelerators based on laser solid interaction have three orders of magnitude higher acceleration gradient in comparison to conventional accelerators.\cite{hegelich2006laser} The obtained proton beams with the characters of ultrashort duration, extreme brightness and small emittance are suitable for a series of potential applications, for instance, production of warm dense matter\cite{bang2015visualization}, injectors for conventional accelerators \cite{busold2015towards} and tumor therapy\cite{bulanov2002oncological}.

The accuracy of target positioning can directly influence the intensity of laser pulse irradiated on the target, which is a key point to improve the cut-off energy of laser-driven protons. The well-studied proton acceleration mechanism named target normal sheath acceleration \cite{wilks2001energetic} (TNSA) predicted a relationship between the maximum proton energy and laser intensity as $E_{max}\propto I^{1/2}$ theoretically.\cite{mora2003plasma} Scaling laws investigated in a huge number of experiments worldwide indicated $E_{max}\propto I^{0.7-1}$ for short duration pulses ($30\sim100$ fs) and $E_{max}\propto I^{0.5}$ for long duration pulses ($0.3\sim 1.0$ ps).\cite{borghesi2008laser,zeil2010scaling} Another important acceleration mechanism promising for the generation of quasi-monoenergetic protons named radiation pressure acceleration (RPA) also predicted a scaling law as $E_{max}\propto I$.\cite{esirkepov2004highly,yan2009self} In order to increase the intensity, and consequently, to improve the maximum energy of the proton beams, small f-number off axis parabolic (OAP) mirrors are widely used to tightly focus the laser beam to micrometer-scale spot. However, a tightly focused pulse will lead to an extreme short Rayleigh length, which requires precise target positioning. A number of experiments have been performed to investigate the dependency of the proton energy on the position of the targets with respect to the laser focus spot.\cite{metzkes2012scintillator,fourmaux2013investigation,wang2016focal} For instance, it's reported that \SI{8}{\mu m} difference on shooting position can result in 30\% difference on the maximum proton energy and nearly one order of magnitude larger proton flux when an f/2 OAP is used.\cite{singh2016diagnostic}

In order to precisely position the target, several methods have been proposed and utilized in laser-driven proton acceleration experiments. A widely used indirect detection method is transverse shadowgraphy imaging, which is suitable for few-micrometer-thick targets by imaging the thin side of the target using a CCD camera.\cite{noaman2017statistical} Retro-focus system has also been applied to target positioning. This scheme injects a green pulse through the back of a dielectric infra-red mirror and then images the back-scattered light from the target.\cite{carroll2011assessment} Another indirect detection method employing multi-wavelength interferometer can realize an accuracy of \SI{10}{\mu m} with the use of a three-wavelength interferometer based on a Mach-Zehnder design. The target is placed at one arm of the interferometer as a reflecting mirror to record depth changes.\cite{booth2014high} However, these indirect detection schemes are not very reliable due to the absence of target surface imaging. Directly imaging of the target surface can be realized by using Questar telephoto microscope \cite{wang2016focal} or a long work distance objective.\cite{singh2016diagnostic} Recently, Gao et al. improved the objective imaging method with a confocal distance sensor and demonstrated the possibility of aligning nanometer targets with 1 Hz repetition rate. \cite{gao2017automated}

\begin{figure*}[b]
\centering
\includegraphics[width=0.9\textwidth]{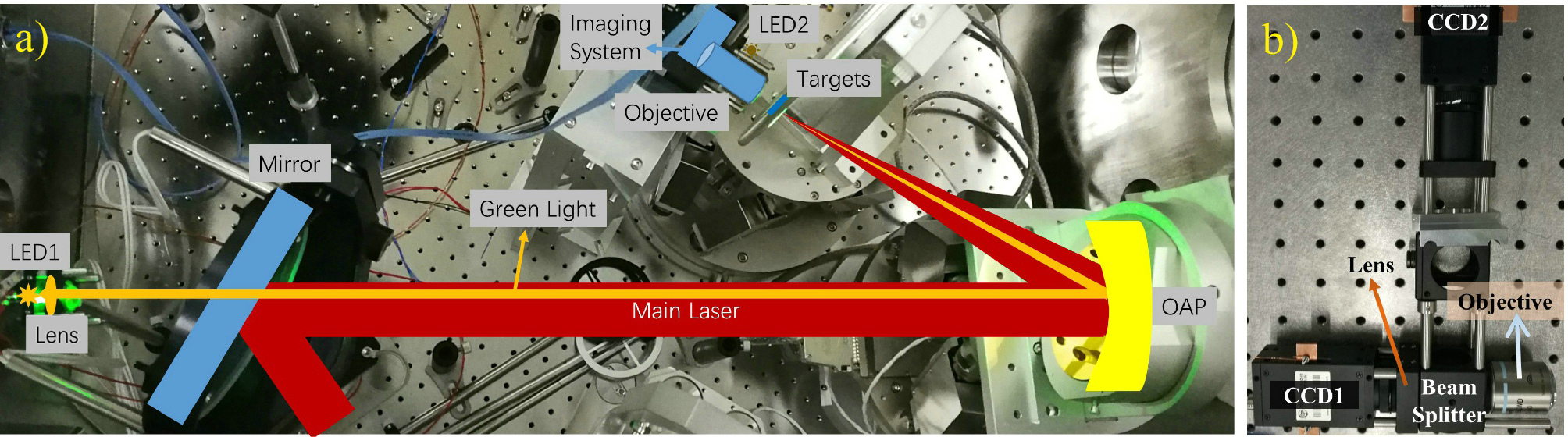}
\caption{Layout of the experimental set up. a) Real picture of the target chamber and positioning system. b) Sketch of the imaging system.}
\label{f1}
\end{figure*}
\section{Target positioning system}
Nowadays, transparent targets such as plastic foils\cite{seuferling2017efficient}, DLC targets\cite{ma2011preparation}, liquid crystal films\cite{poole2014liquid} and $Si_3N_4$ targets\cite{dollar2013high}, are widely applied in laser-driven proton accelerations. Proton beams with cut-off energy higher than \SI{80}{MeV} have been obtained from plastic targets.\cite{wagner2016maximum,kim2016radiation} However, precise positioning of such transparent targets by directly imaging method normally is difficult. One reported solution is putting visible objects, for example, tiny dots or opaque coating patterns, on the surface of transparent targets\cite{poole2014liquid}. But it makes the target production process more complicated. In the present work, we report a positioning system for both opaque and transparent targets with an accuracy of \SI{2}{\mu m}. Defocusing method\cite{agero2003cell,mesquita2006defocusing,roma2014total} is applied to our positioning system to make transparent targets visible as phase objects. Precise positioning of transparent targets can be realized by minimizing the image contrast of the phase objects. Fast positioning by surveying the radius of spherical aberration ring of the phase objects is also realized. The reliability of this positioning system has been verified by performing laser proton acceleration experiments with varied defocusing distances.

The layout of the target positioning system is displayed in Figure \ref{f1}a). The main laser with center wavelength of \SI{800}{nm} is reflected by a dielectric mirror and focused to a spot with full width at half maximum (FWHM) of \SI{5}{\mu m} by an F/3.5 OAP, which corresponds to a Rayleigh length of \SI{40}{\mu m} for a superGaussian laser beam. A green illumination light with center wavelength of \SI{525}{nm} generated from LED1 is collimated by an aspheric condenser lens, and propagates through the dielectric mirror along the laser beam path. The green light then goes through the transparent targets into the imaging system behind the targets. Another collimated white LED source (LED2) is placed above the objective to shed light on the targets for the illumination of the rear surface of opaque targets. The targets are fixed on a special-designed target wheel sitting on a 6-axis Hexapod stage (H-824, PI, repeatability $\pm$\SI{0.5}{\mu m}). The target imaging system mounted on three-axis motorized linear stages consists of a 50-fold long working distance objective (numerical aperture = 0.4), a beam splitter, an achromatic lens and two CCD cameras as shown in Figure \ref{f1}b). 10-fold (CCD1) and 50-fold (CCD2) magnifications can be obtained simultaneously by the use of the beam splitter and an achromatic lens. The 10-fold magnification is suitable for seeking the target center and laser spot due to its larger field of view. The 50-fold camera is beneficial for high precision target positioning and laser spot optimization.

\begin{figure}[ht]
\centering
\includegraphics[width=0.45\textwidth]{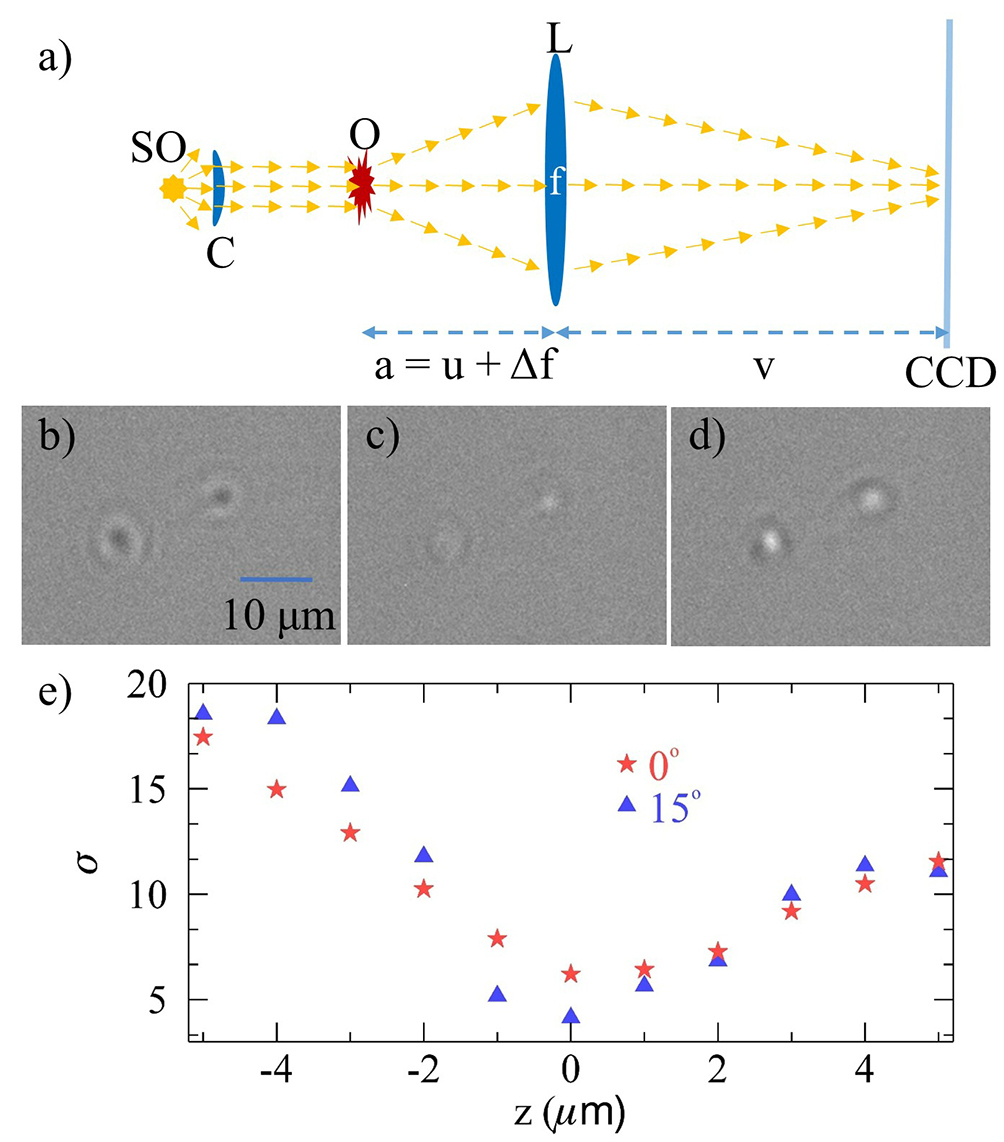}
\caption{a) Sketch of the light path of the imaging system. b), c) and d) are the images of targets at varied defocusing distances \SI{-3}{\mu m}, \SI{0}{\mu m} and \SI{3}{\mu m}, respectively. e) displays the standard deviations of the pixel values of the images versus defocusing distance $\Delta f$ with target tilting angles of $0^\circ$ and $15^\circ$. }
\label{f2}
\end{figure}
\section{Defocusing method}
When transparent targets are close to the image plane of the imaging system, micrometer-scale dots can be observed on the CCDs. The image of these dots varies when the target position is changed. Here we define defocusing distance $\Delta f$ as the distance between the real target position and image plane. Negative $\Delta f$ means that the objective approaches the target, positive $\Delta f$ otherwise. Figure \ref{f2}b), c) and d) show typical images of a plastic target at defocusing distance \SI{-3}{\mu m}, \SI{0}{\mu m} and \SI{3}{\mu m}, respectively. Little center-dark dots are observed when the target is at \SI{-3}{\mu m}. They become almost indistinguishable and turn to center-bright dots as the target move to \SI{0}{\mu m} and \SI{3}{\mu m}, respectively. The evolution of the dots on the targets can be quantitatively characterized by calculating the standard deviations $\sigma$ of the pixel values of the images. Practically, large $\sigma$ corresponds to a high image contrast. Figure \ref{f2}e) displays $\sigma$ of images versus defocusing distance $\Delta f$ with varied target tilting angles $0^\circ$ and $15^\circ$. Here target tilting angle is the incident angle of the laser pulse with respect to the target normal. The standard deviations have a minimum value at $\Delta f = 0$ when the dots almost disappear. Highly precise target positioning can be realized by minimizing $\sigma$ by moving the target step by step.

These dots observed in the images originate from phase objects such as local defects or thickness fluctuations on the transparent target. The emergence and variations of the dots can be interpreted by analysis of the angular spectrums of the transmitted light using the Fourier optics. Figure \ref{f2}a) displays a schematic diagram of the optical path of the imaging system. $SO$ represents the light source, a green LED, $C$ is the condenser lens, $O$ represents the object, a transparent target, $L$ is the objective with an approximate focal length $f$, $u$ and $v$ are conjugated object and image distances. The image distance $v$ is fixed while the actual object distance $a$ is varied from the conjugated object distance $u$ with the defocusing distance $\Delta f$. The propagation of the light through our imaging system can be expressed by the formalism of the angular spectrum.\cite{agero2003cell, goodman2005introduction} The initial angular spectrum out of the target can be expressed as
\begin{equation}\label{e1}
A_0(\vec{q},z)=\int E_0(\vec{\rho},z)e^{-i\vec{q}\cdot\vec{\rho}}d\vec{\rho}
\end{equation}
if we assume the green light propagating along the $z$ direction with angular components in the $x\_y$ plane. Here $\vec{\rho}=x\hat{i}+y\hat{j}$ and $\vec{q}=k_x\hat{i}+k_y\hat{j}$. 
The final angular spectrum on the CCD plane can be expressed as
\begin{equation}\label{e4}
\begin{aligned}
A_3(\vec{q})=&\frac{f}{2\pi ik}e^{ik(a+v)}e^{i[(f-v)/2k]q^2}\\
&\times\int A_0(\vec{\xi})e^{i[(f-a)/2k]\xi^2}e^{-i(f/k)\vec{q}\cdot\vec{\xi}}d\vec{\xi}
\end{aligned}
\end{equation}
To simplify this expression, we make use of the geometrical relationship $a=u+\Delta f$ and imaging formula $f=(u+v)/uv$, to obtain
\begin{equation}\label{e5}
\begin{aligned}
E(\vec{\rho})=&\frac{1}{(2\pi)^2}\frac{f}{2\pi ik}e^{ik(u+v)}\int A_0(\vec{\xi}) d\vec{\xi} \\
&\times\int e^{-\frac{i}{2k}\frac{(v\vec{q}+u\vec{\xi})^2}{u+v}}e^{i\vec{\rho} \cdot \vec{q}}(1-\frac{i\Delta f}{2k}\xi^2)d\vec{q}
\end{aligned}
\end{equation}
Here we use the approximation $\frac{\Delta f}{2k}\xi^2 \ll 1$ for small defocusing distance. Using the integral formula $\int e^{-Aq^2\pm 2B\vec{q}}d\vec{q}=\frac{\pi}{A}e^{\frac{2B^2}{A}}$ and substituting $\vec{q'}=v\vec{q}+u\vec{\xi}$ in Equation \ref{e5} one obtains
\begin{equation}\label{e6}
E(\vec{\rho})=C\frac{1}{(2\pi)^2}\int A_0(\vec{\xi}) e^{-i\frac{u\vec{\xi}\vec{\rho}}{v}} (1-\frac{i\Delta f}{2k}\xi^2)d\vec{\xi}
\end{equation}
where $C=-\frac{u}{v}e^{ik(u+v)}e^{\frac{ik(u+v)}{2v^2}\rho^2}$.
Utilizing the relationship
\begin{equation}\label{e7}
\nabla^2 E_0(\vec{\rho})=\frac{1}{(2\pi)^2}\int A_0(\vec{q}) e^{i\vec{q}\vec{\rho}} (-q^2)d\vec{q}
\end{equation}
we obtain
\begin{equation}\label{e8}
E(\vec{\rho})=C[E_0(-\frac{u}{v}\vec{\rho})+\frac{i\Delta f}{2k}\nabla^2 E_0(-\frac{u}{v}\vec{\rho})]
\end{equation}
By assuming that the target is illuminated by a plane wave, the electric field of this light after passing a phase object can be written as $E_0(\vec{\rho})=E_0e^{i\varphi(\vec{\rho})}$. Finally, the image intensity is
\begin{equation}\label{e9}
I(\vec{\rho})=\frac{u^2}{v^2}E_0^2[1+\frac{\Delta f}{k}\nabla^2 \varphi_0(-\frac{u}{v}\vec{\rho})]
\end{equation}
The phase difference $\varphi(\vec{\rho})$ introduced by a phase object will lead to a change of $I(\vec{\rho})$ when $\Delta f \neq 0$. Different signs of $\Delta f$ can result in center-dark or center-bright dots. So the contrast of the phase objects in the images can reflect the amount of the defocusing distance. The minimal $\sigma$ is obtained at $\Delta f = 0$ as shown in Fig. \ref{f2}e), which can be used for a precise target positioning.

\begin{figure}[ht]
\centering
\includegraphics[width=0.45\textwidth]{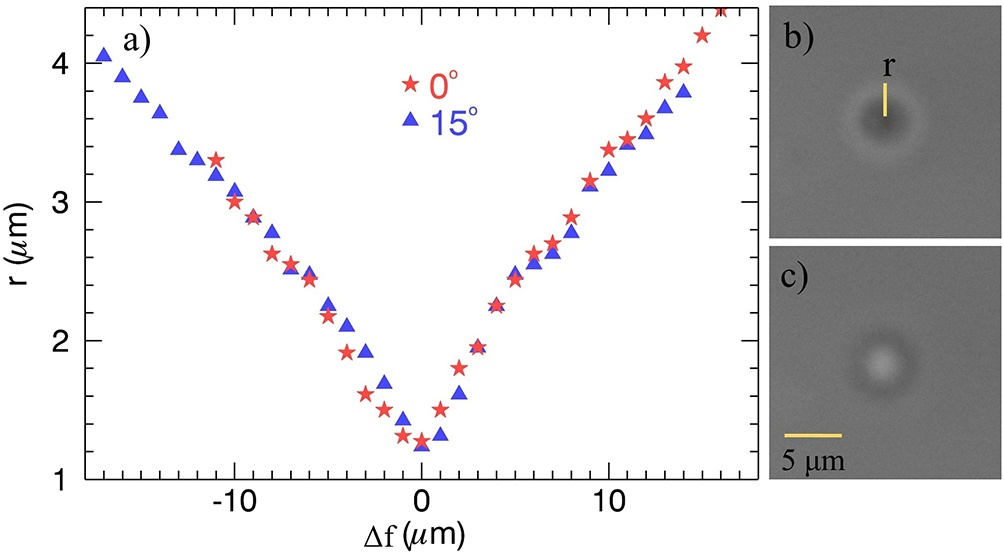}
\caption{a) The ring radius $r$ of a phase object versus defocusing distance $\Delta f$ from experiment with varied target tilting angles $0^\circ$ and $15^\circ$. b) and c) show images of a phase object at different defocusing distances \SI{-8}{\mu m} and \SI{8}{\mu m}. }
\label{f3}
\end{figure}

Another interesting phenomenon we found in the experiments is the linear relationship between the radius of these dots' outer rings and defocusing distance $\Delta f$. Figure \ref{f3}b) and c) display the image of one phase object at different $\Delta f$ \SI{-8}{\mu m} and \SI{8}{\mu m}. One can see the presence of spherical aberration rings of the phase object when the target is out of focus. When $\Delta f$ is large, the approximation $\frac{\Delta f}{2k} \xi^2\ll 1$ will not be met any more. The wave vector of a typical phase object is $\xi=2\pi/R$, here $R\approx$ \SI{3}{\mu m} is the diameter. As a result, it requires $\Delta f \ll$ \SI{6}{\mu m}, which means a tiny defocusing distance. Aberration rings will appear when $\Delta f >$ \SI{6}{\mu m}, which have be utilized in three-dimensional tracking of micron-scale particles\cite{wu2005three,edwards2013near}. Figure \ref{f3}a) displays the ring radius $r$ of defocused images versus defocusing distance $\Delta f$ from experiment with varied target tilting angles $0^\circ$ and $15^\circ$. In the $15^\circ$ case spherical aberration will lead to an ellipse whose semi-major axis corresponding to $r$. The ring radius is linearly proportional to the defocusing distance in a wide range. As a result, one can obtain the defocusing distance quickly based on the ring radius and previous calibration results. This out-of-focus method can be applied to fast targets positioning. The target can be moved to the focus plane directly based on the calculated defocusing distance rather than scanning $\Delta f$ step by step. A target positioning program makes use of Hough transformation has the potential to realize accurate target position in high repetition experiments.

The accuracy of the target positioning system is mainly determined by the objective's depth of focus and the repeatability of the Hexapod (\SI{0.5}{\mu m}). The depth of focus in our imaging system is $\lambda/(2NA^2)=$ \SI{1.7}{\mu m}. It should be noted that the incident angle of the laser makes little influence on the ellipse's semi-major axis of phase objects. Without regard to the laser spot drifting in full energy case, the precision of the target positioning system is within \SI{2}{\mu m}.

\begin{figure}[ht]
\centering
\includegraphics[width=0.45\textwidth]{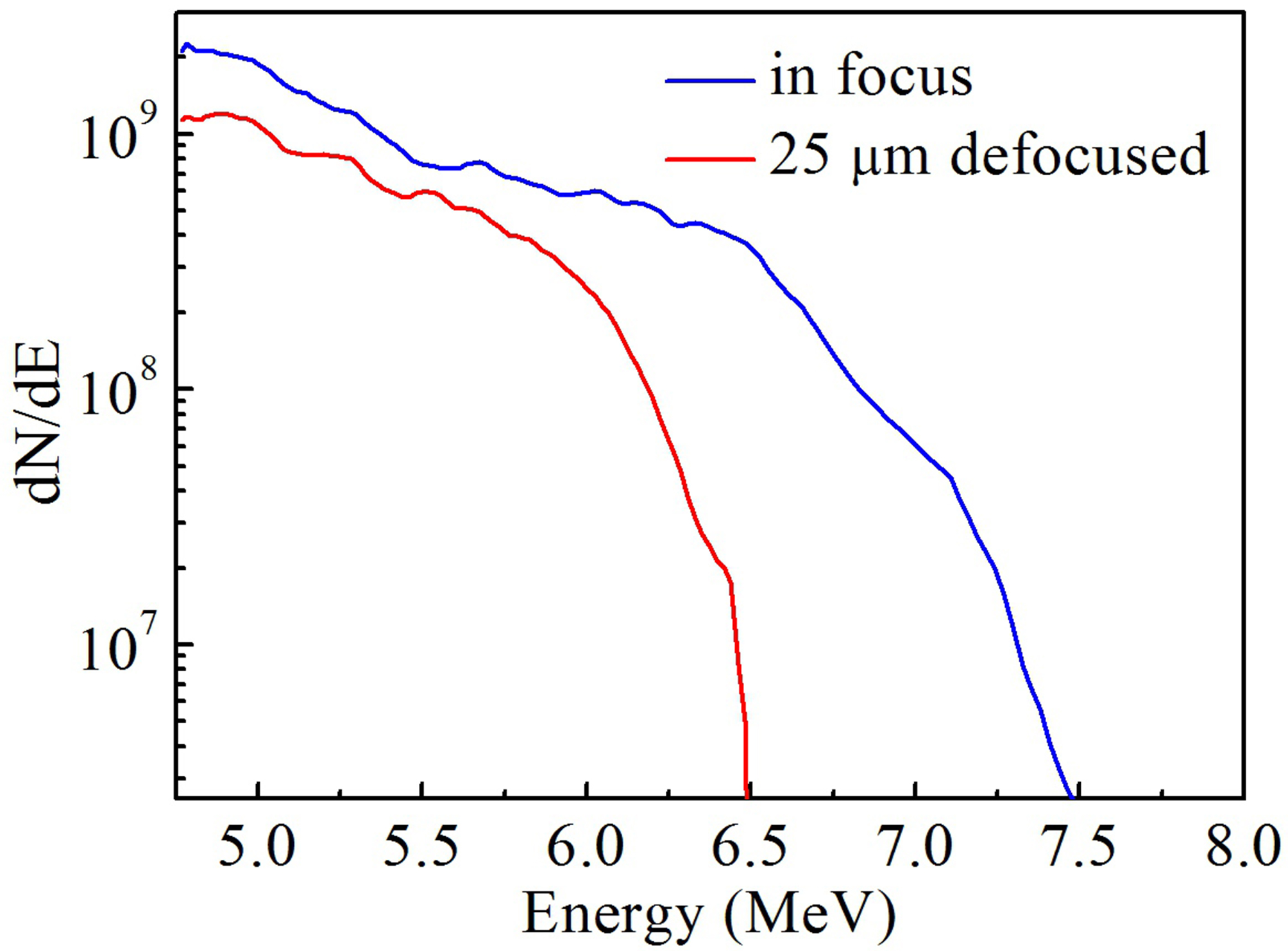}
\caption{The spectrum of protons in focus and \SI{25}{\mu m} defocused from a \SI{1.2}{\mu m} thick plastic target. }
\label{f4}
\end{figure}

\section{Acceleration results}
Laser proton acceleration experiments have been performed to verify the reliability of this positioning system. The experimental arrangement is displayed in Fig. \ref{f1}a). A p-polarized, \SI{30}{fs} laser pulse was irradiated on a \SI{1.2}{\mu m} thick plastic target at an angle of $15^\circ$. The FWHM of the laser spot was \SI{5.9}{\mu m}, corresponding to intensity of \SI{6e19}{W/cm^2} for a \SI{1.6}{J} pulse. The energy spectrum of accelerated protons along the target normal direction was recorded by a Thomson parabola spectrometer and a microchannel plate (MCP) detector. Laser target coupling could be realized by positioning every target and optimizing the laser spot on the same focal plane of the target positioning system. The recorded parabolic traces of protons were analyzed using a Labview code. Figure \ref{f4} shows the spectrum of protons in focus and \SI{25}{\mu m} defocused respectively. Obviously higher cut-off energy and flux of protons can be obtained when the target is in focus. This result justify the necessity of a precise target positioning system.

\section{Conclusions}
We report a positioning system for both opaque and transparent targets with an accuracy of \SI{2}{\mu m} for a compact laser proton accelerator. Precise positioning of transparent targets based on defocusing method can be realized by means of minimizing the image contrast of the phase objects. Fast targets positioning also can be achieved by means of the linear relationship between the spherical aberration ring radius and defocusing distance. Such an accurate positioning system is beneficial for the generations of proton beams with high quality and repeatability.

\section*{Acknowledgements}
This work was supported by National Basic Research Program of China (Grant No. 2013CBA01502), National Natural Science Foundation of China (Grants No. 11575011, 11535001, 61631001, 11775010) and National Grand Instrument Project (2012YQ030142).

\end{document}